\documentclass[preprint,aps]{revtex4}

\usepackage{graphicx}
\usepackage{dcolumn}
\usepackage{bm}
\usepackage{latexsym}
\usepackage{amsfonts}
\usepackage{amssymb}
\usepackage{amsmath}
\usepackage[usenames]{color}

\begin{document}
\preprint{KUNS-2211}

\title{Instability of Small Lovelock Black Holes in Even-dimensions}

\author{Tomohiro Takahashi}
\author{Jiro Soda}
\affiliation{Department of Physics,  Kyoto University, Kyoto, 606-8501, Japan
}

\date{\today}

\begin{abstract}
We study the stability of static black holes in Lovelock theory
which is a natural higher dimensional generalization of Einstein theory.   
We derive a master equation for tensor perturbations in general Lovelock theory.
It turns out that the resultant equation is characterized by one functional 
which determines the background black hole solutions.
Thus, the stability issue of static black holes under tensor perturbations
in general dimensions is reduced to an algebraic problem.
We show that small Lovelock black holes in even-dimensions
are unstable. 
\end{abstract}

\maketitle

\section{Introduction}

It is well known that string theory can be formulated only in ten dimensions. 
 Hence, it is necessary to reconcile this prediction with our real world
 by compactifying extra-dimensions or by considering braneworld.
Intriguingly, in the context of the braneworld with large extra-dimensions,
black holes could be created at the TeV scale~\cite{Giddings:2001bu}.
Hence, the stability of higher dimensional black holes becomes important
since these black holes could be produced at the LHC 
if the spacetime has larger than six dimensions.

The stability of higher dimensional black holes has been 
an active topic since the seminal papers by Kodama and Ishibashi~\cite{Kodama:2003jz}.
It is important to study various black holes in Einstein theory
because black holes produced at the LHC
are expected to be charged or rotating. 
A numerical study of charged black holes has been done~\cite{Konoplya:2007jv}.
To investigate the stability of rotating black holes,
a group theoretical method is developed~\cite{Murata:2007gv}.
The method is used to study the stability of squashed black
 holes~\cite{Kimura:2007cr,Ishihara:2008re}
 and 5-dimensional rotating black holes~\cite{Murata:2008yx}.
The stability of rotating black holes in more than 5-dimensions is also
studied~\cite{Kunduri:2006qa,Oota:2008uj,Kodama:2008rq}.
It is also important to consider the stability of black holes
 in more general gravitational theories
  because black holes are produced at 
the Planck scale where Einstein theory would be no longer valid.
In fact, it is known that Einstein theory is merely a low energy limit of
 string theory~\cite{Boulware:1985wk}. In string theory,
  there are higher curvature corrections in addition to
Einstein-Hilbert term~\cite{Boulware:1985wk}. Thus, it 
is natural to extend gravitational theory into those
 with higher power of curvature in higher dimensions.
  It is Lovelock theory 
that belongs to such class of theories~\cite{Lovelock:1971yv}.
In Lovelock theory, it is known that
there exist static black hole solutions~\cite{Wheeler:1985nh}. 
Hence, it is natural to suppose black holes produced at the LHC are 
of this type~\cite{Barrau:2003tk}.
Thus, it is important to study the stability of these Lovelock black holes.

In the case of second order Lovelock theory, the so-called 
Einstein-Gauss-Bonnet theory, the stability analysis under tensor
perturbations has been performed~\cite{Dotti:2004sh} (see also an earlier work~\cite{Neupane:2003vz}). 
The analysis has been also extended to the scalar and vector 
perturbations~\cite{Gleiser:2005ra}.
It is shown that there exists the scalar mode instability in five dimensions, 
the tensor mode instability in six dimensions, 
and no instability in other dimensions.
In the case of third order Lovelock theory, the stability analysis 
of Lovelock black holes under tensor perturbations 
has been done by us~\cite{Takahashi:2009dz}. 
We have shown that there is the instability for small black holes in eight dimensions.
Although third order Lovelock theory is the most general theory
in seven and eight dimensions, it is not so in more than eight dimensions.
For example, when we consider ten dimensional black holes, 
we need to incorporate fourth order Lovelock terms. 
Indeed, when we consider black holes at the LHC, 
it is important to consider these higher order Lovelock terms~\cite{Rychkov:2004sf}.
Hence, in this paper, we study the stability of black holes in any order
Lovelock theory, namely, in any dimensions. 

The organization of this paper is as follows.
 In Section \ref{seq:2}, we review Lovelock theory and 
explain a graphical method for constructing Lovelock black hole solutions. 
In Section \ref{seq:3}, we present a master equation for tensor perturbations
in the background of Lovelock black holes and reveal its universal structure.
In Section \ref{seq:4}, we examine the stability of Lovelock black holes
 with the method developed previously~\cite{Takahashi:2009dz}. 
 Finally, we summarize our results in Section \ref{seq:5}.

\section{Lovelock Black Holes}
\label{seq:2}

In this section, we review Lovelock theory and introduce
a graphical method to obtain asymptotically flat
black hole solutions. 

In \cite{Lovelock:1971yv}, the most general symmetric, divergence free rank (1,1) tensor 
is constructed out of a metric and its first and second derivatives.
The corresponding Lagrangian can be constructed from $m$-th order Lovelock terms
\begin{eqnarray}
  {\cal L}_m = \frac{1}{2^m} 
  \delta^{\lambda_1 \sigma_1 \cdots \lambda_m \sigma_m}_{\rho_1 \kappa_1 \cdots \rho_m \kappa_m}
  R_{\lambda_1 \sigma_1}{}^{\rho_1 \kappa_1} \cdots  R_{\lambda_m \sigma_m}{}^{\rho_m \kappa_m}
                       \ ,
\end{eqnarray}
where  $R_{\lambda \sigma}{}^{\rho \kappa}$ is the Riemann tensor in $D$-dimensions
and $\delta^{\lambda_1 \sigma_1 \cdots \lambda_m \sigma_m}_{\rho_1 \kappa_1 \cdots \rho_m \kappa_m}$ is the 
generalized totally antisymmetric Kronecker delta. 
Then, Lovelock Lagrangian in  $D$-dimensions is defined by
\begin{eqnarray}
  L = \sum_{m=0}^{k} c_m {\cal L}_m \ ,   \label{eq:lag}
\end{eqnarray}
where we define the maximum order $k\equiv [(D-1)/2]$ and  $c_m$ are 
arbitrary constants. 
Here, $[z]$ represents the maximum integer satisfying $[z]\leq z$. 
Hereafter, we set $c_0=-2\Lambda$, $c_1=1$ and $c_m=a_m/m\ (m\geq 2)$ for convenience. 
Taking variation of the Lagrangian with respect to the metric,
 we can derive Lovelock equation
\begin{eqnarray}
	0={\cal G}_{\mu}^{\nu}
      =\Lambda \delta_{\mu}^{\nu}-\sum_{m=1}^{k}\frac{1}{2^{(m+1)}}\frac{a_m}{m} 
	 \delta^{\nu \lambda_1 \sigma_1 \cdots \lambda_m \sigma_m}_{\mu \rho_1 \kappa_1 \cdots \rho_m \kappa_m}
       R_{\lambda_1 \sigma_1}{}^{\rho_1 \kappa_1} \cdots  R_{\lambda_m \sigma_m}{}^{\rho_m \kappa_m}  \ . \label{eq:EOM}
\end{eqnarray}

As is shown in \cite{Wheeler:1985nh}, there exist static 
exact solutions of Lovelock equation. 
Let us consider the following metric
\begin{eqnarray}
   ds^2=-f(r)dt^2 + \frac{dr^2}{f(r)}+r^2{\bar \gamma}_{i j}dx^idx^j \ ,\label{eq:solution}
\end{eqnarray}
where ${\bar \gamma}_{ij}$ is the metric of $n\equiv D-2$-dimensional constant
 curvature space with a curvature $\kappa$=1,0 or -1. 
Using this metric ansatz, we can calculate Riemann tensor components as
\begin{eqnarray}
	R_{tr}{}^{tr}=-\frac{f^{''}}{2},\ R_{ti}{}^{tj}=R_{ri}{}^{rj}=-\frac{f^{'}}{2r}\delta_{i}{}^{j},\ R_{ij}{}^{kl}=\left(\frac{\kappa-f}{r^2}\right)\left(\delta_{i}{}^{k}\delta_{j}{}^{l}-\delta_{i}{}^{l}\delta_{j}{}^{k}\right) \ . \label{eq:riemann}
\end{eqnarray}
Substituting (\ref{eq:riemann}) into (\ref{eq:EOM}) and defining a new variable
$\psi(r)$ by
\begin{eqnarray}
       f(r)=\kappa-r^2\psi(r) \ , \label{eq:def}
\end{eqnarray}
we obtain an algebraic equation 
\begin{eqnarray}
       W[\psi]\equiv\sum_{m=2}^{k}\left[\frac{a_m}{m}\left\{\prod_{p=1}^{2m-2}(n-p)\right\}\psi^m\right]+\psi-\frac{2\Lambda}{n(n+1)}=\frac{\mu}{r^{n+1}}  \ .
\label{eq:poly}
\end{eqnarray}
In (\ref{eq:poly}), we used $n = D-2$ and $\mu$ is a constant of integration
 which is related to the ADM mass as~\cite{Myers:1988ze}:
\begin{eqnarray}
	M=\frac{2\mu\pi^{(n+1)/2}}{\Gamma((n+1)/2)} \ , \label{eq:ADM}
\end{eqnarray}
where we used a unit $16\pi G=1$.

\begin{figure}[h]
		  \begin{center}
		    \includegraphics[height=8cm, width=12cm]{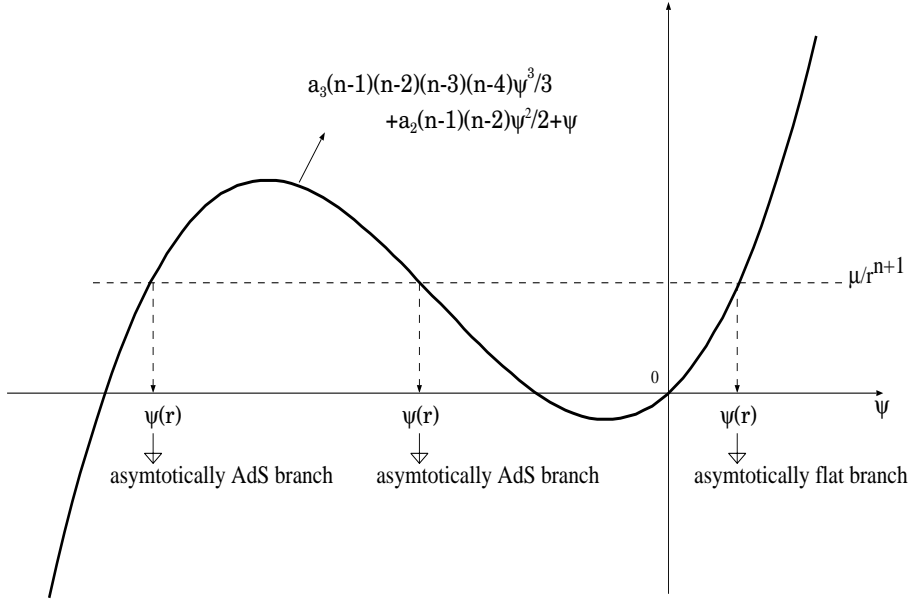}
		    \caption{We illustrate a graphical method for $n=5$ case.
                 In this case, third order Lovelock theory
		             is a complete theory. Therefore, $W[\psi]$ in (\ref{eq:poly})
		 reads a cubic polynomial. 
             In this figure, three roots are depicted. Among these roots, 
		             only $\psi\geq 0$ one corresponds to an
                          asymptotically flat solution.}
		   \label{fig:1} 
		  \end{center} 
	\end{figure}

From (\ref{eq:poly}), it is easy to see that $f(r)$ has many branches.
 In this paper, we want to concentrate on asymptotically flat 
spherically symmetric, $\kappa=1$, solutions 
with a positive ADM mass  $\mu>0$
  because such black holes could be created at the LHC. 
We also assume that Lovelock coefficients satisfy
\begin{eqnarray}
	\Lambda=0,\quad a_m\geq 0\ \label{eq:conditions}
\end{eqnarray}
for simplicity. For example, 
consider $n=5$ for which Eq.(\ref{eq:poly}) becomes third order algebraic equation.   
Though we have a formula for solutions of Eq.(\ref{eq:poly}) in this case,
 the roots are too complicated in general. 
Hence, we use a graphical method illustrated in Fig.\ref{fig:1}. 
Because of the conditions (\ref{eq:conditions}),
the function is monotonic for positive $\psi$. 
From (\ref{eq:poly}), we see the root behaves $\psi \sim \mu/r^{n+1}$ 
or $f(r)\sim 1-\mu/r^{n-1}$ as $r\rightarrow \infty$.
Thus, the asymptotically flat solutions belong to the branch where $\psi$ 
is always positive. 
 
\begin{figure}[h]
		  \begin{center}
		    \includegraphics[height=8cm, width=12cm]{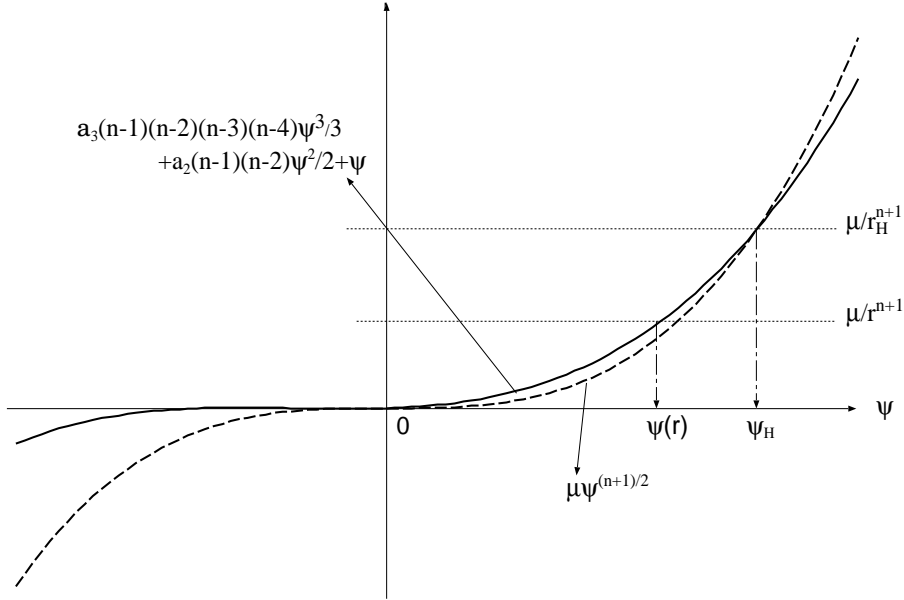}
		    \caption{For $n=5$ case, this figure explains a
                method for calculating $\psi_H$ or $r_H$ graphically.}
		   \label{fig:2} 
		  \end{center} 
	\end{figure}

Note that the horizon 
radius of the asymptotically flat solution is characterized  by $f(r_H)=0$.
 From (\ref{eq:def}), we have a relation $\psi_H\equiv \psi(r_H)=1/r_H^2$.
  Using this relation and (\ref{eq:poly}), we obtain an algebraic equation
\begin{eqnarray}
              W[\psi_H]  =\mu \psi_H^{(n+1)/2} \ .
 \label{eq:rh}
\end{eqnarray}	
This determines $\psi_H$ and hence $r_H$. In Fig.\ref{fig:2}, 
we present a graphical method to solve Eq.(\ref{eq:rh}). 
 It is obvious from Fig.\ref{fig:2} that $\infty\geq r \geq r_H$ corresponds to
$0\leq \psi \leq \psi_H$ when $f(r)$ describes an asymptotically flat solution. 
It is also obvious that $\psi_H$ becomes larger as $\mu$ is smaller.  Note that
there may be no horizon if $\mu$ is too small (for example, in seven dimensions, there is no horizon if $0\leq\mu\leq 8a_3 -9a_2^2$).

Using the metric ansatz (\ref{eq:solution}), the Kretschmann scalar $R_{a b c d}R^{a b c d}$ is 
\begin{eqnarray}
	R_{a b c d}R^{a b c d}=f^{''}+2n\frac{f^{'2}}{r^2}+2n(n-1)\frac{(\kappa-f)^2}{r^4}.\nonumber
\end{eqnarray}
Then, this solution has curvature singularities at $r=0$
 or where derivatives of $f(r)$ diverges.  
For asymptotically flat spherical symmetric solutions, 
the conditions $|1-f|<\infty$, $|f^{'}|<\infty$ and $|f^{''}|<\infty$ are satisfied 
except for $r=0$. Therefore, 
there is a curvature singularity only at $r=0$ in the cases we are considering.
 And asymptotically flat solutions have a horizon 
if the parameter $\mu$ is sufficiently large. 
So, this solution doesn't have a naked singularity and 
describes a black hole with a mass M defined in (\ref{eq:ADM}) and no angular momentum.

\section{master equation for tensor perturbations}
\label{seq:3}

To analyze the stability, it is convenient to decompose the metric 
under the symmetry of $n$-dimensional symmetric space.
There are scalar, vector, and tensor modes.
In this paper, we study tensor perturbations. 
In this section, we do not restrict parameters.

We consider tensor perturbations around the solution (\ref{eq:solution})
\begin{eqnarray}
	\delta g_{a b}=0 \ , \quad \delta g_{a i}=0 \ ,\quad 
      \delta g_{i j}=r^2 \phi(t,r)\bar{h}_{i j}(x^i ) \ , 
\end{eqnarray}
where $a,b=(t,r)$ and $\phi (t,r)$ represents the dynamical degrees of freedom.
Here, $\bar{h}_{ij}$  are defined by
\begin{eqnarray}
	\bar{\nabla}^{k}\bar{\nabla}_{k} \bar{h}_{ij}=\gamma \bar{h}_{ij} \ , \qquad
	\bar{\nabla}^{i} \bar{h}_{ij}=0 \ ,\quad \bar{\gamma}^{ij}\bar{h}_{ij}=0.
\end{eqnarray}
Here, $\bar{\nabla}^{i}$ denotes a covariant derivative with respect to
 $\bar{\gamma}_{ij}$. Here, the eigenvalue is given by 
$\gamma =-\ell (\ell +n-1)+2$, ($\ell =2,3,4 \cdots$) for $\kappa=1$ 
and  negative real number for $\kappa=-1,0$.

Perturbing the Lovelock equation (\ref{eq:EOM}), we can get the first order perturbation equation
\begin{eqnarray}
	0=\delta{\cal G}_{\mu}^{\nu}=-\sum_{m=1}^{k}\frac{a_m}{2^{(m+1)}} 
	 \delta^{\nu \lambda_1 \sigma_1 \cdots \lambda_m \sigma_m}_{\mu \rho_1 \kappa_1 \cdots \rho_m \kappa_m}
       R_{\lambda_1 \sigma_1}{}^{\rho_1 \kappa_1} \cdots  R_{\lambda_{m-1} \sigma_{m-1}}{}^{\rho_{m-1} \kappa_{m-1}} \delta R_{\lambda_m \sigma_m}{}^{\rho_m \kappa_m},  \label{eq:pert}
\end{eqnarray}
where $\delta R_{a b}{}^{c d}$ represents the first order variation of the Riemann tensor. 
Since we are considering tensor perturbations, apparently we have
$\delta{\cal G}_{a}{}^{b} = \delta{\cal G}_{a}{}^{i}=0$.
To calculate $\delta {\cal G}_{i}^{j}$, we only need
 $\delta R_{ti}{}^{tj}$, $\delta R_{ri}{}^{rj}$ and $\delta R_{ij}{}^{kl}$. 
 The reason is simple.
We consider, for example, the term proportional to $\delta R_{tk}{}^{lm}$ 
in $\delta{\cal G}_{i}{}^{j}$. 
This means $\lambda_m=t$ or $\sigma_m=t$ in (\ref{eq:pert}); and $\rho_m \neq t$ and $\kappa_m \neq t$. 
Since $\delta_{\rho_1\kappa_1\cdots}^{\lambda_1 \sigma_1\cdots}$ consists of Kronecker delta,
 there must exist 
$\ell\ (1\leq \ell \leq m-1)$ such that $\rho_\ell =t$ or $\kappa_\ell =t$. Then $\lambda_\ell$ or $\sigma_\ell$ must 
be $t$ because of background Riemann tensor (\ref{eq:riemann}). So, for example, $\lambda_m$ and $\lambda_\ell$ become $t$. 
Noticing that $\delta_{\rho_1\kappa_1\cdots}^{\lambda_1 \sigma_1\cdots}$ is totally
 antisymmetric, we see $\delta_{\rho_1\kappa_1\cdots}^{\lambda_1 \sigma_1\cdots}=0$ 
when $\lambda_m = \lambda_\ell$.
 Thus, the coefficient of $\delta R_{tk}{}^{lm}$ in $\delta{\cal G}_{i}{}^{j}$ must be 0. 
For the same reason, coefficients of other than 
$\delta R_{ti}{}^{tj}$, $\delta R_{ri}{}^{rj}$ and $\delta R_{ij}{}^{kl}$ vanish. 
From the metric ansatz (\ref{eq:solution}), the necessary components can be deduced as
\begin{eqnarray}
	\delta R_{ti}{}^{tj}
      &=&\left(\frac{{\ddot \phi}}{2f}-\frac{f^{'}\phi^{'}}{4}\right) {\bar h}_{i}{}^{j}
      \ , \nonumber\\ 
      \delta R_{ri}{}^{rj}&=&\left(-\frac{f\phi^{''}}{2}+\left(-\frac{f^{'}}{4}-\frac{f}{r}\right)\phi^{'}\right) {\bar h}_{i}{}^{j} \ , \nonumber\\
      \delta^m_l \delta R_{im}{}^{jl}&=& 
      \left[ -\frac{n-2}{2} \frac{f}{r} \phi' 
           +\frac{2\kappa -\gamma}{2r^2} \phi \right]{\bar h}_{i}{}^{j}
      \ .\label{eq:deltariemann}
\end{eqnarray}
Thus, using the following relation
\begin{eqnarray}
	\delta_{ik_1k_2\cdots k_{m-1}k_{m}}^{jl_1l_2\cdots l_{m-1}l_{m}}\delta_{l_{m}}^{k_{m}}=(n-m)\delta_{ik_1k_2\cdots k_{m-1}}^{jl_1l_2\cdots l_{m-1}} \ , \label{eq:formulae}
\end{eqnarray}
we can calculate $\delta {\cal G}_{i}^{j}$ as follows: 
\begin{eqnarray}
	\delta{\cal G}_{i}{}^{j}&=&\frac{h(r)}{r^{n-2}}\left(\delta R_{ti}{}^{tj}+\delta R_{ri}{}^{rj}\right)+\frac{h^{'}}{(n-2)r^{n-3}}\delta^m_l \delta R_{im}{}^{jl}\nonumber\\
	                      &=&\frac{1}{r^{n-2}}\left[\frac{h}{2f}\left({\ddot \phi}-f^2\phi^{''}\right)-\left(\frac{f^{'}h}{2}+\frac{fh}{r}+\frac{fh^{'}}{2}\right)\phi^{'}+\frac{(2\kappa-\gamma)h^{'}}{2(n-2)r}\phi\right],\label{eq:deltaG}
\end{eqnarray}
where 
\begin{eqnarray}
	h(r)&=&r^{n-2}+\sum_{m=2}\Biggl[a_mr^{n-2m}(\kappa-f)^{m-2}\left(\prod_{p=2}^{2m-2}(n-p)\right)\nonumber\\
	&\ &\hspace{3cm}\times\left\{-(m-1)rf^{'}+(n-(2m-1))(\kappa-f)\right\}\Biggr]. \label{eq:h_function}
\end{eqnarray}
It is useful to note that the universal function $h$ is related to $W$ as
\begin{eqnarray}
  h(r) = \frac{d}{dr}\left[ \frac{r^{n-1}}{n-1}  \frac{dW[\psi]}{d\psi} \right] \ .
  \label{h:W}
\end{eqnarray}

Separating the variables $\phi(r,t)=\chi(r)e^{\omega t}$, 
we can deduce the master equation
from (\ref{eq:pert}) and (\ref{eq:deltaG}) as follows: 
\begin{eqnarray}
	- f^2\chi^{''}
      - \left( f^2 \frac{h^{'}}{h}+ \frac{2f^2}{r}
      +  f f^{'} \right) \chi^{'}
      +  \frac{(2\kappa-\gamma)f}{(n-2)r}\frac{h^{'}}{h} \chi = -  \omega^2 \chi \ .
 \label{eq:master_eq}
\end{eqnarray}
Here, we should notice that
we have not used information about Lovelock coefficients and $f(r)$. 
Hence, this master equation itself is very general, although we will consider 
specific situations in the next section. 
Moreover, this equation is the same as the master equation 
in~\cite{Takahashi:2009dz} except for 
the concrete form of $h(r)$. Hence, the method used in~\cite{Takahashi:2009dz}
 can also be applicable.

\section{Even-Dimensional Small Black Holes are Unstable}
\label{seq:4}

In this section, we first clarify condition for the stability
of black holes. Then, we demonstrate that small black holes are
always unstable in even-dimensions. 

\subsection{Condition for Stability}

 In this section, we assume the conditions (\ref{eq:conditions}) are satisfied.
  Then, there exists an asymptotically flat branch. 
  And we also assume spherical symmetry and positivity of the mass,
   i.e., $\kappa=1$ and $\mu>0$.
   
As we will soon see, the master equation
(\ref{eq:master_eq}) can be transformed into the Schr${\rm {\ddot o}}$dinger form. 
To do this, we have to impose the condition 
\begin{eqnarray}
	h(r)>0 \ ,  \quad ({\rm for}\ r>r_H) \ . \label{eq:assum}
\end{eqnarray}
In fact, this is necessary for the linear analysis to be applicable. 
 In the case that there exists $r_0$ such that $h(r_0)=0$ and $r_0>r_H$,
 we encounter a singularity. 
In fact, using approximations $h(r)\sim h' (r_0)(r-r_0)\equiv h'(r_0) y$, 
$f(r)= f(r_0 ) $ and $r=r_0$, we can show
the solution near $r_0$ is given by $\chi \sim c_1 + c_2 \log y$, 
where $c_1$ and $c_2$ are constants of integration. 
The solution is singular at $y =0$ for generic perturbations. 
 The similar situation occurs even in cosmology with higher derivative
  terms~\cite{Kawai:1998bn,Satoh:2007gn}. 
 In those cases, this kind of singularity alludes to ghosts.
 Indeed, there is a region $h(r)<0$ outside horizon where kinetic
 term of perturbations has wrong sign.
 Hence, we simply assume these black hole solutions are not allowed. 

When the condition (\ref{eq:assum}) is fulfilled, introducing a new variable
$
	\Psi(r)=\chi(r)r\sqrt{h(r)} 
$
 and switching to the coordinate $r^*$, defined by $dr^*/dr=1/f$,
we can rewrite Eq.(\ref{eq:master_eq}) as
\begin{eqnarray}
	-\frac{d^2\Psi}{dr^{*2}}+V(r(r^*))\Psi=-\omega^2\Psi\equiv E\Psi \ , 
      \label{eq:shradinger}
\end{eqnarray}
where
\begin{eqnarray}
	V(r)=\frac{(2\kappa-\gamma)f}{(n-2)r}\frac{d \ln{h}}{dr}+\frac{1}{r\sqrt{h}}f\frac{d}{dr}\left(f\frac{d}{dr}r\sqrt{h}\right) \label{eq:potential}
\end{eqnarray}
is an effective potential.

For discussing the stability, the "S-deformation" approach
 is useful~\cite{Kodama:2003jz, Dotti:2004sh}.  Let us define the operator 
\begin{eqnarray}
	{\cal H}\equiv -\frac{d^2}{dr^{*2}}+V
\end{eqnarray}
acting on smooth functions defined on $I=(r^{*}_H,\infty)$.
Then, (\ref{eq:shradinger}) is the eigenequation and $E$ is eigenvalue of ${\cal H}$.
We also define the inner products as
\begin{eqnarray}
	(\varphi_1,\varphi_2)=\int_I \varphi_1^*\varphi_2 dr^*\ .
\end{eqnarray}
In this case, for any $\varphi$, we can find a smooth function $S$ such that 
\begin{eqnarray}
	(\varphi,{\cal H}\varphi)=\int_{I} (|D\varphi|^2+\tilde{V}|\varphi|^2)dr^{*},
\end{eqnarray}
where we have defined
\begin{eqnarray}
	D=\frac{d}{dr^{*}}+S  \ , \quad 
	\tilde{V}=V+f\frac{dS}{dr}-S^2  \ .
\end{eqnarray}
Following~\cite{Dotti:2004sh}, we choose $S$ to be
\begin{eqnarray}
	S=-f\frac{d}{dr}\ln{(r\sqrt{h})} \ .
\end{eqnarray}
Then, we obtain the formula
\begin{eqnarray}
	(\varphi,{\cal H}\varphi)
      =\int_{I} |D\varphi|^2dr^{*}+(2\kappa-\gamma)
      \int_{r_H}^{\infty}\frac{|\varphi|^2}{(n-2)r}\frac{d \ln{h}}{dr}dr \ .
       \label{eq:stab}
\end{eqnarray}
Here, the point is that the second term in (\ref{eq:stab}) includes
a factor $2\kappa-\gamma >0$, but $h$ does not include $\gamma$. 
Hence, by taking a sufficiently large $2\kappa -\gamma$, we can 
always make the second term dominant.  

Now, let us show that the sign of $d\ln h /dr$ determines the stability.
If $d\ln h /dr >0$ on $I$, the solution (\ref{eq:solution}) is stable.
This can be understood as follows. 
Note that $2\kappa-\gamma>0$, 
then we have $\tilde{V}>0$ for this case.
 That means $(\varphi,{\cal H}\varphi)>0$ for arbitrary $\varphi$ if $d \ln{h}/dr>0$ on $I$. 
We choose, for example, $\varphi$ as the lowest eigenstate, then we can conclude that the lowest eigenvalue $E_0$ is positive. 
Thus, we proved the stability. The other way around,
if $d \ln{h} /dr <0$ at some point in $I$, the solution is unstable. 
To prove this, the inequality
\begin{eqnarray}
	\frac{(\varphi,{\cal H}\varphi)}{(\varphi,\varphi)} \geq E_0 
      \label{eq:ineq}
\end{eqnarray}
is useful. This inequality is correct for arbitrary $\varphi$.
 If $d \ln{h}/dr <0$ at some point in $I$, we can find $\varphi$ such that
\begin{eqnarray}
	 \int_{r_H}^{\infty}\frac{|\varphi|^2}{(n-2)r}\frac{d \ln{h}}{dr}dr<0 \ .
\end{eqnarray}
In this case, (\ref{eq:stab}) is negative for sufficiently large $2\kappa-\gamma$.
 Then, the inequality (\ref{eq:ineq}) implies $E_0<0$ and the solution has unstable modes. 
Thus, we can conclude that the solution is stable if and only if
 $d \ln{h}/dr>0$ on $I$.
 
To summarize, we need to check the sign of $dh/dr$ outside the horizon $r>r_H$
 to investigate the stability. 
Note that the condition $h(r) >0$ is necessary to have healthy black holes.

\subsection{Instability of Small Black Holes}

Here, we assume $h(r)$ is positive outside the horizon.
Otherwise, such black holes are already sick. Hence, what we have to
do is to check the sign of $dh/dr$. 

In order to see the sign of $dh/dr$, it is very useful to express it  
as function of $\psi$ instead of that of $r$. Using (\ref{eq:poly}) and its derivative,
we obtain 
\begin{eqnarray}
	\left(\partial_{\psi} W[\psi]\right) \psi^{'}=-(n+1)\frac{\mu}{r^{n+1}}=-(n+1)\frac{W[\psi]}{r} \ .
\end{eqnarray}
The above formula can be used to eliminate $\psi'$ in (\ref{h:W}).
The result reads
\begin{eqnarray}
	h(r)=r^{n-2}\left(\partial_{\psi}W-\frac{n+1}{n-1}\frac{W\partial_{\psi}^2W}{\partial_{\psi}W}\right) \ .
\end{eqnarray}
Similarly, $dh/dr$ can be written as
\begin{eqnarray}
	\frac{dh}{dr}&=&r^{n-3}\Biggl[(n-2)\partial_{\psi}W-\frac{(n+1)(n-4)}{n-1}\frac{W\partial_{\psi}^2W}{\partial_{\psi}W}\nonumber\\
	&\ &\hspace{1.5cm}+\frac{(n+1)^2}{n-1}\frac{W^2\left\{\partial_\psi W \partial_\psi^3 W-(\partial_\psi^2 W)^2\right\}}{(\partial_\psi W)^3}\Biggr]\ . 
\end{eqnarray}
Since $W[\psi]$ is a polynomial function of $\psi$ and $\partial_\psi W$
is positive, we can determine the sign of $dh/dr$ by examining
the sign of polynomial determined by the above formula. 
Thus, the stability problem has been reduced to an algebraic one.

Substituting the explicit form of $W[\psi]$ into this equation, 
we obtain
\begin{eqnarray}
	\frac{dh}{dr}=r^{n-3}\frac{L[\psi]}{\left(1+\sum_{m=2}^{k}\left[a_m\left\{\prod_{p=1}^{2m-2}(n-p)\right\}\psi^{m}\right]\right)^3}\ . \label{eq:dhpsi}
\end{eqnarray}
Here, the lowest and leading term of $L(x)$ is 
\begin{eqnarray}
	L(x)&=&(n-2)+\cdots\nonumber\\
	&\ &\ +\frac{a_k^4}{k^2}(n-1)^3\left\{\prod_{p=2}^{2k-2}(n-p)^4\right\}(n-(2k-1))(n-(3k-1))x^{4k-4} \ . \label{eq:L}
\end{eqnarray}
We note that the highest order $k=[(D-1)/2]$ is related to
the dimensions as $n=2k-1$ in odd-dimensions and $n= 2k$ in even-dimensions.
In odd-dimensions, the leading term disappears.
Hence, we cannot say anything in general. Hence, we consider only
even-dimensions.

Let us examine the sign of $L(x)\ (x\geq0)$. 
If $n=2k$, the coefficient of the lowest term is positive and 
that of the leading one of (\ref{eq:L}) becomes negative. 
Therefore $L(x)>0$ near $x=0$ and $L(x)<0$ for large $x$. This means that there exists 
roots of $L(x)=0$ because $L(x)$ is continuity function; 
we define $x_0$ as the lowest positive root. 
If $\psi_H<x_0$, then $L[\psi]>0$ for $0\leq\psi\leq \psi_H$,
 and hence we see $dh/dr>0$ for $r>r_H$. 
While, if $\psi_H>x_0$, then there exists a region $L[\psi]<0$ in $x_0\leq\psi\leq \psi_H$,
 and thus there exists an area $dh/dr<0$ in $r>r_H$. 
Therefore, black holes are stable 
if $\psi_H < x_0$ and unstable if $\psi_H>x_0$. 
Considering the result that $\psi_H$ becomes larger 
as $\mu$, which is related to ADM mass, becomes smaller,
we conclude that there exist critical mass below which black holes become 
unstable.

Summarizing this section, there is critical mass below which black holes are
 unstable in $n=2k$. Thus, we have shown the instability of small Lovelock black holes
 in even-dimensions.

\section{Conclusion}
\label{seq:5}

We have studied the stability of static black holes in Lovelock theory
which is a natural higher dimensional generalization of Einstein theory.   
We derived a master equation for tensor perturbations in general Lovelock theory.
It turned out that the resultant equation is characterized by one functional $W[\psi]$
which determines the background black hole solutions through Eq.(\ref{eq:poly}).
The stability issue of static black holes under tensor perturbations
in general dimensions has been reduced to an algebraic problem.
We have shown that there exists the instability of Lovelock black holes 
with small masses in even-dimensions. Remarkably, the instability
is strong on short distance scales. Curiously, the similar instability also appears
in the Gauss-Bonnet cosmology~\cite{Kawai:1998ab}.

We have done the analysis of tensor perturbations and found the instability
of small black holes in even-dimensions.
Taking into account the results in Gauss-Bonnet theory~\cite{Gleiser:2005ra},
if we extend the analysis to scalar and vector perturbations,
we can expect that small black holes are unstable also in odd-dimensions.
If so, we need more profound understanding of the instability in Lovelock theory.

It is also interesting to investigate the fate of the instability.
As the instability is stronger for higher multipole orders $\ell$, 
the resultant geometry would be weird. 
This issue is very important because black holes lose their mass due to
the Hawking radiation and eventually become unstable. 

Related to the above, we have to find meaning of the universal function $h(r)$. 
As was shown in this paper, this function governs 
 the stability of black holes. 
Therefore, if $h(r)$ has, for example, thermodynamical meaning, the relation between 
thermodynamical~\cite{Dehghani:2005vh} and dynamical instability might be revealed. 

\begin{acknowledgements}
This work is supported by the Japan-U.K. Research Cooperative Program, 
Grant-in-Aid for Scientific Research Fund of the Ministry of 
Education, Science and Culture of Japan No.18540262
and by the Grant-in-Aid for the Global COE Program 
"The Next Generation of Physics, Spun from Universality and Emergence". 
\end{acknowledgements}

\end{document}